\documentstyle[12pt]{article}
\begin{document}
\baselineskip=14pt
\begin{titlepage}
\vspace*{0.0cm}

\begin{center}
{\bf  Amplitude equations for coupled electrostatic waves\\
in the limit of weak instability}
\end{center}
\begin{center}
John David Crawford \\
Department of Physics and Astronomy\\
University of Pittsburgh\\
Pittsburgh, PA 15260
\end{center}
\begin{center}
Edgar Knobloch\\
Department of Physics\\
University of California\\
Berkeley, CA 94720
\end{center}
\vspace{0.5cm}

\centerline{\bf ABSTRACT}

\begin{quote}
We consider the simplest instabilities involving multiple unstable
electrostatic plasma waves corresponding to four-dimensional systems of mode
amplitude equations.  In each case the coupled amplitude equations are
derived up to third order terms. The nonlinear coefficients are singular in
the limit in which the linear growth rates vanish together.   These
singularities are analyzed using techniques developed in previous studies
of a single unstable wave. In addition to
the singularities familiar from the one mode problem, there are new
singularities
in coefficients coupling the modes.
The new singularities are most severe when the
two waves have the same linear phase velocity and satisfy the spatial
resonance condition $k_2=2k_1$. As a result the short wave mode saturates at
a dramatically smaller amplitude than that predicted for the weak growth
rate regime on the basis of single mode theory. In contrast the long wave mode
retains the single mode scaling. If these resonance conditions are not
satisfied both modes retain their single mode scaling and saturate at
comparable
amplitudes.
\end{quote}

\vfill

\begin{center}
August 21, 1997
\end{center}
\end{titlepage}
\tableofcontents
\baselineskip=24pt

\section{Introduction}

Recently we presented a detailed analysis of the amplitude equation for
a single unstable electrostatic mode in an unmagnetized Vlasov plasma
(henceforth (I)).\cite{jdcaj97} The analysis reveals a fundamental difficulty
with the derivation of amplitude equations for this class of problems: the
coefficients in the amplitude equations become singular in the limit in which
the growth rate $\gamma$ of the unstable wave is allowed to vanish. Although
these singularities can be removed by an appropriate $\gamma$-dependent
rescaling (see below) the analysis shows that amplitude equations of this type
cannot be truncated at any finite order.\cite{jdcaj96,jdc95}
Nonetheless the scaling identified by
the theory predicts the amplitude at which weakly growing waves will saturate,
and hence is of fundamental importance both in plasma physics and in the
closely related problem of shear flow instability of ideal fluids.

In view of the importance of the predicted scaling for applications we
investigate here the effects of including additional unstable modes. We
consider only
the simplest possibilities, those requiring a four-dimensional system of
amplitude equations. There are three such instabilities distinguished by
the symmetry of the equilibrium and whether the unstable modes have real or
complex eigenvalues. We find that unless the two modes satisfy a strong
resonance condition the presence of the second mode does not alter the
saturation amplitude of the original mode. However, in the important resonant
case in which the phase velocities of the two modes are the same and their
wavenumbers $k_1$, $k_2$ satisfy $k_2=2k_1$ the long wave mode suppresses
dramatically the saturation amplitude of the short wave mode.

The amplitude equation defines, in the limit $\gamma\rightarrow0^+$, a kind
of singular perturbation problem whose detailed features reveal asymptotic
scaling behavior of the nonlinear wave. This is a key idea behind our approach
and the reader is referred to (I) for a more detailed discussion. For the
single mode instabilities there is only one amplitude $A(t)$ and one
seeks a scaling $A(t)=\gamma^\beta a(\gamma t)$ such that the evolution
equation for $a(\tau)$, $\tau\equiv\gamma t$,
has a nonsingular limit as $\gamma\rightarrow 0^+$. In dissipative
problems, the critical eigenvalues are isolated on the imaginary axis, and
$\beta=1/2$ is the generically expected exponent. As a result, in the generic
case, the amplitude equation can be truncated at third order. This is not
so for an unstable electrostatic wave. Here the situation is quite different
because the Vlasov equation is
Hamiltonian and the eigenvalues of the mode merge with a continuous spectrum
on the imaginary axis at criticality, i.e. as $\gamma\rightarrow 0^+$. As a
consequence
the nonlinear coefficients in the amplitude equation are singular as
$\gamma\rightarrow0^+$ signalling  strong nonlinear effects
that saturate the unstable linear growth at exceptionally small
wave amplitudes. A quantitative signature of this reduction of the nonlinear
wave amplitude is a larger exponent: $\beta=5/2$ for plasmas with multiple
mobile species and $\beta=2$ in the limiting case of infinitely massive (fixed)
ions and mobile electrons.
In fact, the analysis presented in (I) showed that
setting $\beta=5/2$ yielded a theory that was finite to all orders in the
amplitude expansion as
$\gamma\rightarrow0^+$.

In this paper we investigate the coupled amplitude equations for two
unstable modes with amplitudes $A(t)$ and $B(t)$. The coupled equations
contain the single mode instabilities due to excitation of only $A$ or only
$B$,
and the previously studied singular coefficients govern these special
cases.  There are now separate scalings possible for each amplitude,
\begin{equation}
|A(t)|=\gamma_1^{\beta_1} a(\gamma_1 t) \hspace{1.0in}
|B(t)|=\gamma_2^{\beta_2} b(\gamma_2 t),\label{eq:scalings}
\end{equation}
and we know $\beta_j\geq5/2$, $j=1,2$
is required to control the singularities in the single mode coefficients.
We seek to determine the possible singularities in the coupling coefficients
between $A$ and $B$ that are new; specifically we wish to know if these
singularities can dominate the single mode singularities and require new
nonlinear scalings for the instability with two simultaneously growing
waves.

In the remainder of this introduction, we summarize our notation and
state some relevant results about the linearized theory. Section
\ref{sec:expand} enumerates the possible instabilities described by
four-dimensional systems of amplitude equations. In each case the general
form of the amplitude equations can be anticipated on the basis of symmetry.
Section \ref{subsec:exp} describes the procedure for calculating the
coefficients in these amplitude equations
and summarizes the results for the leading
terms through third order. The singularities that arise in these terms
in the limit of weak instability are analyzed in Section
\ref{sec:sing2m}, and their consequences are discussed in Section
4. The paper ends with a brief conclusion.

\subsection{Notation}
We briefly summarize the notation of (I) which will be used.
Our model is the one-dimensional, multi-species Vlasov plasma defined
by
\begin{equation}
\frac{\partial F^{(s)}}{\partial t}+v\frac{\partial F^{(s)}}{\partial x}+
\kappa^{(s)}\;E\;\frac{\partial F^{(s)}}{\partial v}=0,
\hspace{0.3in}
\frac{\partial E}{\partial
x}=\sum_s\int^\infty_{-\infty}\,dv\,F^{(s)}(x,v,t).\label{eq:vlasov}
\end{equation}
Here $x$, $t$ and $v$ are measured in units of $u/\omega_e$, $\omega_e^{-1}$
and $u$, respectively, where $u$ is a chosen velocity scale and
$\omega_e^2=4\pi e^2n_e/m_e$.
The plasma length is $L$ with periodic boundary conditions and
\begin{equation}
\int^{L/2}_{-L/2}\,dx\,\int^\infty_{-\infty}\,dv\,F^{(s)}(x,v,t)=
\left(\frac{z_s\;n_s}{n_e}\right)L,\label{eq:Fnorm}
\end{equation}
where $q_s=e\,z_s$ is the charge of species $s$ and $\kappa^{(s)}\equiv
{q_sm_e/em_s}$. Note that $\kappa^{(e)}=-1$ for electrons and
that the normalization (\ref{eq:Fnorm}) for negative species
makes the distribution function negative.

Let $F_0(v)$ and $f(x,v,t)$ denote the multi-component fields for the
equilibrium and
perturbation, respectively, and $\kappa$ the matrix of mass ratios,
\begin{equation}
 f\equiv\left(\begin{array}{c}
f^{(s_1)}\\f^{(s_2)}\\ \vdots\end{array}\right)\;\;\;\;
F_0\equiv\left(\begin{array}{c}F_0^{(s_1)}\\F_0^{(s_2)}\\
\vdots\end{array}\right)\;\;\;\;
\kappa\equiv\left(\begin{array}{cccc}\kappa^{(s_1)} & 0 & 0 & \cdots\\
 0&\kappa^{(s_2)}&0&\cdots\\ \vdots&\vdots&\vdots\end{array}\right),
\end{equation}
then the system (\ref{eq:vlasov}) can be concisely
expressed as
\begin{equation}
\frac{\partial f}{\partial t}={\cal{L}}\,f+{\cal{N}}(f)\label{eq:dynsys}
\end{equation}
where
\begin{eqnarray}
{\cal{L}}\,f&=&\sum^{\infty}_{l=-\infty}\,e^{ilx}\,(L_l f_l)(v)
\label{eq:fexp}\\
(L_l f_l)(v)&=&\left\{\begin{array}{cc}0&l=0\\
-il\left[vf_l(v)+\kappa\cdot\eta_l(v)\sum_{s'}
\int^\infty_{-\infty}\,dv'\,f^{(s')}_l(v')
\right]&l\neq0,\end{array}\right.\label{eq:linop}
\end{eqnarray}
with $\eta_l(v)\equiv-\partial_vF_0/l^2$,
and
\begin{equation}
{\cal{N}}(f)=\sum^{\infty}_{m=-\infty}\,
e^{imx}\,{\sum^{\infty}_{l=-\infty}}'\, \frac{i}{l}\left(\kappa\cdot
\frac{\partial
f_{m-l}}{\partial v}\right)
\sum_{s'}\int^\infty_{-\infty}\,dv'\,f^{(s')}_l(v').
\label{eq:nop}
\end{equation}
In the spatial Fourier expansion (\ref{eq:fexp}), $l$ denotes an integer
multiple of the basic wavenumber $2\pi/L$,  and a primed summation as in
(\ref{eq:nop}) omits the $l=0$ term. The notation $\kappa\cdot\eta_l(v)$ or
$\kappa\cdot\partial_v f_{m-l}$
denotes matrix multiplication.

An inner product is needed in Section \ref{sec:expand} to derive the amplitude
equation. For two multi-component fields of $(x,v)$, e.g.
$B=(B^{(s_1)},B^{(s_2)},B^{(s_3)},..)$ and
$D=(D^{(s_1)},D^{(s_2)},D^{(s_3)},..)$, we define their inner product by
\begin{eqnarray}
(B,D)&\equiv&\sum_s\int^{L/2}_{-L/2}\,dx\int_{-\infty}^{\infty}\,dv\,
B^{(s)}(x,v)^\ast D^{(s)}(x,v)
=\int^{L/2}_{-L/2}\,dx\,<B,D>,
\end{eqnarray}
where
\begin{eqnarray}
<B,D>&\equiv&\sum_s \int_{-\infty}^{\infty}\,dv\,B^{(s)}(x,v)^\ast
D^{(s)}(x,v).
\end{eqnarray}

\subsection{Summary of linear theory}

The spectral theory for ${\cal{L}}$ is well established, and the needed results
are simply recalled to establish our notation.\cite{vkamp,case,cra1} The
eigenvalues $\lambda\equiv-il z$ of ${\cal{L}}$ are determined by the roots
$\Lambda_{l}(z)=0$ of the ``spectral function'',
\begin{equation}
\Lambda_{l}(z)\equiv 1+
\int^\infty_{-\infty}\,dv\,\frac{\sum_s\kappa^{(s)}\eta_l^{(s)}(v)}{v-z}.
\label{eq:specfcn}
\end{equation}
The linear dielectric function $\epsilon_{{l}}(z)$ is obtained on replacing
the contour in (\ref{eq:specfcn}) by the Landau contour for ${\rm Im}(z)<0$;
for ${\rm Im}(z)>0$, $\Lambda_{l}(z)$ and $\epsilon_{{l}}(z)$ are the same
function.

Associated with an eigenvalue $\lambda\equiv-il z$ is the multi-component
eigenfunction $\Psi(x,v)=e^{ilx}\,\psi(v)$, where
\begin{equation}
\psi(v)=-\frac{\kappa\cdot \eta_l}{v-z}.
\end{equation}
There is also an associated adjoint eigenfunction
$\tilde{\Psi}(x,v)=e^{ilx}\tilde{\psi}(v)/L$
satisfying $(\tilde{\Psi},\Psi)=1$ with
\begin{equation}
\tilde{\psi}(v)=- \frac{1}{\Lambda'_{l}(z)^\ast(v-z^\ast)}.
\label{eq:aefcn2}
\end{equation}
Note that all components of $\tilde{\psi}(v)$ are the same.
The normalization in (\ref{eq:aefcn2}) assumes that the root of
$\Lambda_{l}(z)$ is simple and is chosen so that $<\tilde{\psi},\psi>=1$.
The adjoint determines the projection of $f(x,v,t)$ onto the eigenvector, and
this projection defines the time-dependent amplitude of $\Psi$, i.e.
$A(t)\equiv(\tilde{\Psi},f)$.

In section \ref{subsec:exp}
some of our results are conveniently stated in terms of
the resolvent operator,
$R_l(w)\equiv(w-L_l)^{-1}$, whose general form
follows from (\ref{eq:linop}) by solving
$(w-L_l)f=g$ for $f$.\cite{jdcaj97,cra1}  Here both $g(v)=(g^{(s_1)}(v),
g^{(s_2)}(v),\ldots)$
and $f$ are multi-component fields, and  $R_l(w)$
acts by
\begin{equation}
R_l(w)\,g=\left(
\begin{array}{c}(R_l(w)\,g)^{(s_1)}(v)\\(R_l(w)\,g)^{(s_2)}(v)\\\vdots
\end{array}\right),
\end{equation}
where
\begin{equation}
(R_l(w)\,g)^{(s)}(v)=\frac{1}{il(v-iw/l)}
\left[g^{(s)}(v)-\frac{\kappa^{(s)}\eta^{(s)}_l}{\Lambda_{l}(iw/l)}
\sum_{s'}\int^{\infty}_{-\infty}
\,dv'\,\frac{g^{(s')}(v')}{v'-iw/l}\right].\label{eq:resol}
\end{equation}


\section{Amplitude equations: general features}\label{sec:expand}

Each of our instabilities can be formulated within a general framework as
follows. The wavenumbers of the unstable modes $\Psi_2(x,v)$ and $\Psi_1(x,v)$
are given by $|k_2|\geq |k_1|>0$, and the corresponding eigenvalues are
$\lambda_j=-ik_jz_j$ for $j=1,2$, where $\Lambda_{k_j}(z_j)=0$.
With periodic boundary conditions, each wavenumber is a
multiple of the minimum $k$, i.e. $k_j=2\pi n_j/L$ with integer $n_j$.
We assume all
roots are simple, i.e. $\Lambda'_{k_j}(z_j)\neq0$; in addition, in the limit
${\rm Im}(z)\rightarrow0$ of weak growth rates, the equation for the root
is given by
\begin{equation}
\sum_s\kappa^{(s)}\eta_{k_j}^{(s)}(v_j)=0,\hspace{0.5in}
1+{\rm P}\int^\infty_{-\infty}\,dv\,
\frac{\sum_s\kappa^{(s)}\eta_{k_j}^{(s)}(v)}{v-v_j}=0,\label{eq:critroot}
\end{equation}
where $v_j={\rm Re}(z_j)$ is the phase velocity at criticality. These
relations  will be important for our analysis of the singularities in
the nonlinear coefficients.

The four-dimensional eigenspace $E^u$ is spanned by
$\{\Psi_1, \Psi_1^\ast,\Psi_2, \Psi_2^\ast\}$, and
the components of the distribution function
along the unstable eigenvectors are identified by
writing
\begin{equation}
f(x,v,t)=\left[A(t)\Psi_1(x,v) + B(t)\Psi_2(x,v) +cc\right] +
S(x,v,t),\label{eq:linmodes}
\end{equation}
where $A(t)=(\tilde{\Psi}_1,f)$ and $B(t)=(\tilde{\Psi}_2,f)$ are the mode
amplitudes and
$(\tilde{\Psi}_j,S)=0$.
In (\ref{eq:linmodes}), $\tilde{\Psi}_j=\exp(ik_jx)\,\tilde{\psi}_j/L$
is the appropriate adjoint function for $z_j$ from (\ref{eq:aefcn2}).

In the  $(A,B,S)$ variables, the Vlasov equation (\ref{eq:dynsys}) becomes:
\begin{eqnarray}
\dot{A}&=&\lambda_1\,A+(\tilde{\Psi}_1,{\cal{N}}(f))\label{eq:adot}\\
\dot{B}&=&\lambda_2\,B+(\tilde{\Psi}_2,{\cal{N}}(f))\label{eq:bdot}\\
\frac{\partial S}{\partial t}&=&{\cal{L}}
S+{\cal{N}}(f)-\left[(\tilde{\Psi}_1,{\cal{N}}(f))\,\Psi_1 +
(\tilde{\Psi}_2,{\cal{N}}(f))\,\Psi_2 +
cc\right],\label{eq:Sdot}
\end{eqnarray}
where
\begin{equation}
(\tilde{\Psi}_j,{\cal{N}}(f))=-{i}\,{\sum^{\infty}_{l=-\infty}}'
\frac{1}{l}\;
<{\partial_v}\,\tilde{\psi}_j,  \kappa\cdot  f_{k_j-l}>
\sum_{s'}\;\int^\infty_{-\infty}\,dv'\,f^{(s')}_l(v').\label{eq:projnl}
\end{equation}
In writing (\ref{eq:adot}) we have used the adjoint relationship
$(\tilde{\Psi}_j,{\cal{L}} S)=
({\cal{L}}^\dagger\tilde{\Psi}_j,S)=\lambda_j(\tilde{\Psi}_j,S)=0$ and in
(\ref{eq:projnl}) an integration by parts shifts the velocity
derivative onto $\tilde{\psi}$.

These coupled equations are equivalent to
(\ref{eq:dynsys}); however, by restricting them to the unstable manifold we
obtain autonomous equations for $A(t)$ and $B(t)$. The details of this
restriction are discussed in an earlier paper\cite{jdc95} and also in (I).
The unstable manifold is tangent to $E^u$ at the equilibrium, and
near $F_0$ it can be described by a function $H(x,v,A,A^\ast,B,B^\ast)$. Thus
\begin{equation}
f^u(x,v,t)=\left[A(t)\Psi_1(x,v) + B(t)\Psi_2(x,v)  + cc\right] +
H(x,v,A(t),A^\ast(t),B(t),B^\ast(t))\label{eq:fu}
\end{equation}
represents a distribution function on $W^u$.
In this expression, the evolution of $S$ is determined from $H$, i.e.
\begin{equation}
S(x,v,t)=H(x,v,A(t),A^\ast(t),B(t),B^\ast(t))=\left(
\begin{array}{c}H^{(s_1)}(x,v,A,A^\ast,B,B^\ast)\\
H^{(s_2)}(x,v,A,A^\ast,B,B^\ast)\\
\vdots
 \end{array}\right).\label{eq:Su}
\end{equation}
When this representation is
substituted into (\ref{eq:adot}) - (\ref{eq:Sdot}) we obtain
\begin{eqnarray}
\dot{A}&=&\lambda_1\,A+(\tilde{\Psi}_1,{\cal{N}}(f^u))\label{eq:aeqn}\\
\dot{B}&=&\lambda_2\,B+(\tilde{\Psi}_2,{\cal{N}}(f^u))\label{eq:beqn}\\
\left.\frac{\partial S}{\partial t}\right|_{f^u} &=&{\cal{L}}
H+{\cal{N}}(f^u)-\left[(\tilde{\Psi}_1,{\cal{N}}(f^u))\,\Psi_1 +
(\tilde{\Psi}_2,{\cal{N}}(f^u))\,\Psi_2 +
cc\right].\label{eq:Sdotwu}
\end{eqnarray}
Note that equations (\ref{eq:aeqn}) - (\ref{eq:beqn}) define an
{\em autonomous} flow describing
the self-consistent nonlinear evolution of the unstable modes; this is the
four-dimensional system we study.

Symmetries of the Vlasov-Poisson system (\ref{eq:dynsys}) and the
equilibrium $F_0(v)$ are important qualitative features of the problem.
Spatial translation,
${\cal T}_a:(x,v)\rightarrow (x+a,v),$
and reflection,
${{\cal R}:(x,v)\rightarrow (-x,-v)},$
act as operators on $f(x,v,t)$ in the usual way: if $\alpha$
denotes an arbitrary transformation then $(\alpha\;f)(x,v)\equiv
f(\alpha^{-1}\,(x,v))$, where $(\alpha\;f)(x,v)$
denotes the transformed distribution
function. The operators ${\cal{L}}$ and ${\cal{N}}$ commute
with ${\cal T}_a$ due to the spatial homogeneity of $F_0$,
and if $F_0(v)=F_0(-v)$,
then ${\cal{L}}$ and ${\cal{N}}$ also commute with the reflection operator
${\cal R}$.
With periodic boundary conditions, $x$ is a periodic coordinate so
${\cal T}_a$ and ${\cal R}$ generate ${\rm O}(2)$, the symmetry group of the
circle. If only the translation symmetry is present, then the group is ${\rm
SO}(2)$.

The action of
translation ${\cal T}_a$ on $f(x,v,t)$
implies an action on the variables $(A,B,S)$.  From
(\ref{eq:linmodes}) we note that ${\cal T}_af(x,v,t)=f(x-a,v,t)$
is equivalent to
\begin{equation}
{\cal T}_a \,(A,B,S(x,v))=(e^{-ik_1 a}A, e^{-ik_2 a}B,
S(x-a,v)).\label{eq:Atrans}
\end{equation}
The representation of ${\cal R}$ in the variables $(A,B,S)$
depends on specific
details of $\Psi_1$ and $\Psi_2$ in the individual cases discussed below.
These symmetries determine the general form of the amplitude equations
(\ref{eq:aeqn}) - (\ref{eq:beqn}) for each of the instabilities we consider.

\subsection{Instability without reflection symmetry: $F_0(v)\neq F_0(-v)$}
\label{subsec:complex}
When $F_0(v)$ lacks reflection symmetry, the generic four-dimensional
problem arises for modes with unequal wavenumbers $k_2>k_1>0$ and
complex eigenvalues.
The roots, $\Lambda_{k_j}(z_j)=0$,
determine the phase velocities $v_j=\omega_j/k_j$
and the growth rates $\gamma_j$  of the linear modes from the real and
imaginary
parts of the eigenvalue $\lambda_j=-ik_jz_j\equiv\gamma_j-i\omega_j$.
The corresponding
eigenvectors are
\begin{equation}
\Psi_j(x,v)=e^{ik_jx}\,\psi_j(v)=
e^{ik_jx}\left(-\frac{\kappa\cdot \eta_{k_j}}{v-z_j}\right),\;\;j=1,2.
\label{eq:lefcn}
\end{equation}

In this case, the identities
$\Lambda_{k_j}(z_j)=\Lambda_{-k_j}(z_j)$ and
$\Lambda_{k_j}(z_j)^\ast=\Lambda_{k_j}(z_j^\ast)$
imply three additional eigenvectors: $\Psi_j^\ast$, $\Phi_j$ and $\Phi_j^\ast$,
where
\begin{equation}
\Phi_j(x,v)=e^{ik_jx}\,\psi_j(v)^\ast.
\end{equation}
These eigenfunctions correspond to eigenvalues $\lambda_j^\ast$,
$-\lambda_j^\ast$
and $-\lambda_j$, respectively, and fill out the eigenvalue quartets
characteristic of Hamiltonian systems. In the absence of reflection symmetry,
the eigenvalues are typically simple and the four-dimensional unstable subspace
is
spanned by $\{\Psi_1,\Psi_1^\ast,\Psi_2,\Psi_2^\ast\}$. A beam-plasma system
with a weak beam is the prototypical example of this
instability.

Since $F_0(v)$ is spatially homogeneous,  the amplitude equations
(\ref{eq:aeqn}) - (\ref{eq:beqn})
always have translation symmetry
(\ref{eq:Atrans}) and we can apply standard results on the form of such
symmetric equations.~\cite{gss,jdcek}
Hence we know the right hand side of (\ref{eq:aeqn}) - (\ref{eq:beqn})
takes the general form,
\begin{eqnarray}
\dot{A}&=&r(\sigma)\,A+s(\sigma)\,(A^\ast)^{n_2-1}B^{n_1}\label{eq:adef}\\
\dot{B}&=&p(\sigma)\,B+q(\sigma)\,A^{n_2}(B^\ast)^{n_1-1},\label{eq:bdef}
\end{eqnarray}
where $\sigma=(\sigma_1,\sigma_2,\sigma_3,\sigma_4)$ denotes the four basic
invariants
\begin{equation}
\sigma_1=|A|^2\hspace{0.25in}\sigma_2=|B|^2\hspace{0.25in}
\sigma_3=A^{n_2}(B^\ast)^{n_1}\hspace{0.25in}\sigma_4=(A^\ast)^{n_2}B^{n_1}.
\label{eq:invariants}
\end{equation}
The integers $n_j$ refer to the wavenumbers $k_j=2\pi n_j/L$.\cite{relprime}
The complex-valued functions $r,s,p,$ and $q$  are determined from
(\ref{eq:aeqn}) - (\ref{eq:beqn}), but they can
depend on the amplitudes only through the four
invariants $\sigma_1,\sigma_2,\sigma_3$ and $\sigma_4$.

\subsection{Instability with reflection symmetry: $F_0(v)= F_0(-v)$}

When the equilibrium is reflection-symmetric the transformation
${\cal{R}}f(x,v,t)=f(-x,-v,t)$ commutes with ${\cal L}$ and ${\cal{N}}$
in the Vlasov equation. In this setting,
the roots of $\Lambda_{l}(z)$ can be either imaginary or complex, depending on
the detailed form of $F_0(v)$,
and correspondingly we may encounter instabilities due to either real or
complex eigenvalues.

\subsubsection{Real eigenvalues}\label{subsubsec:real}

The description for real eigenvalues is quite similar to the examples without
symmetry and we use the same
notation (\ref{eq:lefcn})
for the unstable eigenvectors $\Psi_1(x,v)$ and $\Psi_2(x,v)$. In this case,
however, both
linear phase velocities are zero. In addition, each of the real eigenvalues
$\lambda_1$ and $\lambda_2$
has multiplicity
two since the states ${\cal{R}}\Psi_1 =\Psi_1^\ast$ and ${\cal{R}}\Psi_2
=\Psi_2^\ast$ are also eigenvectors.
The translation symmetry (\ref{eq:Atrans}) still holds, as well as
the reflection transformation given by
\begin{equation}
{\cal{R}}\, (A,B,S(x,v))=(A^\ast,B^\ast,S(-x,-v)).\label{eq:Aref}
\end{equation}
The form of the amplitude equations (\ref{eq:aeqn}) - (\ref{eq:beqn})
is the same except that the functions $r, s, p$ and $q$
in (\ref{eq:adef}) - (\ref{eq:bdef})
are now real-valued and depend on only three invariants $\sigma_1$, $\sigma_2$,
and
$\sigma_+\equiv\sigma_3+\sigma_4$. This model applies for example to a
reflection-symmetric
two-stream distribution with instability at two wavenumbers.

\subsubsection{Complex eigenvalues}\label{subsubsec:complex}

An instability in a reflection-symmetric system with complex
eigenvalues likewise yields eigenvalues of multiplicity two. If only
one wavelength is involved these result in a four-dimensional problem,
and provide an additional example of what may be called a one-mode
instability, cf.~(I). Such an instability arises, for example, in
a beam-plasma system with counter-propagating beams.\cite{demeio}

In the notation of our general framework, we let $k_1=-k_2=k>0$
and $z_1=-z_2=z_0$; then $\Lambda_{k}(z_0)=0$
implies $\Lambda_{-k}(-z_0)=0$, and
these roots correspond to reflection-related eigenvectors for the eigenvalue
$\lambda=-ikz_0$, i.e. $\lambda=-ik_1z_1=-ik_2z_2$:
\begin{eqnarray}
\Psi_1(x,v)&=&e^{ikx}\,\psi_1(v)=
e^{ikx}\left(-\frac{\kappa\cdot \eta_{k}(v)}{v-z_0}\right)\\
&&\nonumber\\
\Psi_2(x,v)&=&({\cal{R}}\Psi_1)(x,v)=e^{-ikx}\,\psi_1(-v)=
e^{-ikx}\left(-\frac{\kappa\cdot \eta_{k}(v)}{v+z_0}\right).
\end{eqnarray}
These solutions describe oppositely propagating linear waves with phase
velocities $v_1=-v_2=\omega/k$, where $z_0=v_1+i\gamma/k$. In contrast to the
previous examples, where both $z_1$ and $z_2$ sit in the upper-half plane
(corresponding to positive wavenumbers), in this case $z_2$ falls in the
lower-half plane.

Now the mode amplitudes in (\ref{eq:linmodes})
transform according to (\ref{eq:Atrans}) and applying ${\cal R}$ to
(\ref{eq:linmodes}) yields
\begin{eqnarray}
{\cal{R}}\, (A,B,S(x,v))&=&(B,A,S(-x,-v)).\label{eq:hopfref}
\end{eqnarray}
The O(2) symmetry generated by $\cal{R}$ and ${\cal T}_a$ implies amplitude
equations of the form
\begin{equation}
\left(\begin{array}{c}\dot{A}\\ \dot{B}\end{array}\right)=
P(\mu_1,\mu_2,\mu_2^\ast)\left(\begin{array}{c}{A}\\{B}\end{array}\right)
+Q(\mu_1,\mu_2,\mu_2^\ast)
\left(\begin{array}{c}{B^\ast}\\{A^\ast}\end{array}\right),
\label{eq:o2hopf1}
\end{equation}
where $P$ and $Q$ are functions of the invariants $\mu_1\equiv|A|^2+|B|^2$
and $\mu_2=AB$ with $P(0,0,0)=\lambda$ and $Q(0,0,0)=0$. In this case a
further simplification is possible;
terms that do not commute with the ``phase-shift'' symmetry,
$(A,B)\rightarrow(e^{i\phi}A,e^{i\phi}B)$,
can be removed by near-identity coordinate changes, $(A,B)\equiv
(A'+\Phi_1(A',B'), B'+\Phi_2(A',B'))$, to obtain the normal form:
\begin{equation}
\left(\begin{array}{c}\dot{A}\\ \dot{B}\end{array}\right)=
R(\mu_1,\mu_3)\left(\begin{array}{c}{A}\\{B}\end{array}\right)
+S(\mu_1,\mu_3)[|A|^2-|B|^2]\left(\begin{array}{c}{A}\\{-B}\end{array}\right)
\label{eq:o2hopf}
\end{equation}
where $\mu_3\equiv(|A|^2-|B|^2)^2$ and we have dropped the primes
on $(A',B')$.\cite{gss,jdcek}

\subsection{Amplitude expansions}\label{subsec:exp}
We wish to study the nonlinear terms (\ref{eq:projnl}) in the amplitude
equations
(\ref{eq:aeqn}) - (\ref{eq:beqn}). The Fourier components $f_{l}$
follow from (\ref{eq:fu})
\begin{equation}
f^u_l(v)=A\psi_1(v)\delta_{l,k_1}+A^\ast\psi_1(v)^\ast\delta_{l,-k_1}
+B\psi_2(v)\delta_{l,k_2}+B^\ast\psi_2(v)^\ast\delta_{l,-k_2}+H_l(v),
\label{eq:fucoeff}
\end{equation}
and the amplitude expansion of $H_l(v)$ begins with second-order terms,
\begin{eqnarray}
H_l(v)&=&[h_1(v)|A|^2+h_2(v)|B|^2]\,\delta_{l,0}
+h_3(v)BA^\ast\delta_{l,k_2-k_1}+h_3(v)^\ast B^\ast A\delta_{l,k_1-k_2}
\label{eq:gexp}\\
&&+h_4(v)A^2\delta_{l,2k_1}+h_4(v)^\ast(A^\ast)^2\delta_{l,-2k_1}
+h_5(v)AB\delta_{l,k_2+k_1}+h_5(v)^\ast A^\ast B^\ast\delta_{l,-k_2-k_1}
\nonumber\\
&&\hspace{0.3in}+h_6(v)B^2\delta_{l,2k_2}+h_6(v)^\ast(B^\ast)^2\delta_{l,-2k_2}
+{\cal O}(3).\nonumber
\end{eqnarray}
Thus the nonlinear terms (\ref{eq:projnl}) can be written out
in terms of the coefficients
$h_i(v)$ in (\ref{eq:gexp}),
neglecting terms that
are higher than third order in the amplitudes:
\begin{eqnarray}
\lefteqn{(\tilde{\Psi}_1,{\cal{N}}(f^u))=r_{1}(0)A|A|^2+r_2(0)A|B|^2+
\delta_{k_2,2k_1}s(0) A^\ast B+\delta_{k_2,3k_1}s(0) (A^\ast)^2 B}
\label{eq:Aexp}\\
&&+ \delta_{k_2,-k_1}\left[P_2(0)A^2B+
(Q_1(0)+P_3(0))B^\ast|A|^2+Q_3(0)A^\ast
(B^\ast)^2+Q_1(0)B^\ast|B|^2\right]\nonumber
\end{eqnarray}
with Taylor coefficients
\begin{eqnarray}
r_1(0)&=&-\frac{i}{k_1}\left[<\partial_v\tilde{\psi}_1,\kappa\cdot(h_1-h_4)>
+\frac{\Gamma_4}{2}<\partial_v\tilde{\psi}_1,\kappa\cdot{\psi}_1^\ast>\right]\\
&&\nonumber\\
r_2(0)&=&-i\left[\frac{<\partial_v\tilde{\psi}_1,\kappa\cdot h_2>}{k_1}+
\frac{<\partial_v\tilde{\psi}_1,\kappa\cdot(h_3^\ast-h_5)>}{k_2}
-\frac{\Gamma_3^\ast}{k_2-k_1}<\partial_v\tilde{\psi}_1,\kappa\cdot\psi_2>
\right.\nonumber\\
&&\hspace{2.5in}\left.
+\frac{\Gamma_5}{k_1+k_2}
<\partial_v\tilde{\psi}_1,\kappa\cdot\psi_2^\ast>\right]
\label{eq:r20}\\
&&\nonumber\\
s(0)&=&\left\{\begin{array}{lc}
-i\left[{<\partial_v\tilde{\psi}_1,\kappa\cdot{\psi}_1^\ast>}/{k_2}
-{<\partial_v\tilde{\psi}_1,\kappa\cdot{\psi}_2>}/{k_1}\right]&
\hspace{0.0in}{\rm if}\hspace{0.15in}k_2=2k_1\\
&\\
-i\left[{<\partial_v\tilde{\psi}_1,\kappa\cdot h_4^\ast>}/{k_2}
-{<\partial_v\tilde{\psi}_1,\kappa\cdot h_3>}/{k_1}\right.&\\
\hspace{0.5in}
+\Gamma_3<\partial_v\tilde{\psi}_1,\kappa\cdot{\psi}_1^\ast>/(k_2-k_1)
&{\rm if}\hspace{0.15in}k_2=3k_1\\
\left.\hspace{1.5in}
-\Gamma_4^\ast<\partial_v\tilde{\psi}_1,\kappa\cdot{\psi}_2>/2k_1\right]
\end{array}\right.
\label{eq:s0}\\
&&\nonumber\\
P_2(0)&=&-i\left[\frac{<\partial_v\tilde{\psi}_1,\kappa\cdot h_5>}{k_1}+
\frac{<\partial_v\tilde{\psi}_1,\kappa\cdot h_4>}{k_2}
+\frac{\Gamma_4}{2k_1}<\partial_v\tilde{\psi}_1,\kappa\cdot\psi_2>\right]\\
Q_1(0)+P_3(0)&=&
-i\left[\frac{<\partial_v\tilde{\psi}_1,\kappa\cdot (h_5^\ast-h_3^\ast)>}{k_1}
-\frac{<\partial_v\tilde{\psi}_1,\kappa\cdot h_1>}{k_2}
\right.\nonumber\\
&&\hspace{2.5in}\left.
+\frac{\Gamma_3^\ast<\partial_v\tilde{\psi}_1,\kappa\cdot\psi_1^\ast>}{k_1-k_2}
\right]
\label{eq:QP20}\\
&&\nonumber\\
Q_3(0)&=&i\left[\frac{<\partial_v\tilde{\psi}_1,\kappa\cdot
(h_6^\ast-h_5^\ast)>}{k_1}+\frac{\Gamma_6^\ast}{2k_2}
<\partial_v\tilde{\psi}_1,\kappa\cdot{\psi}_1^\ast>\right]\\
Q_1(0)&=&-i\left[
\frac{<\partial_v\tilde{\psi}_1,\kappa\cdot (h_6^\ast- h_2)>}{k_2}
-\frac{\Gamma_6^\ast}{2k_2}
<\partial_v\tilde{\psi}_1,\kappa\cdot{\psi}_2>\right]
\end{eqnarray}
where $\Gamma_j\equiv\sum_s\int dv\,h_j^{(s)}(v)$, and
\begin{eqnarray}
\lefteqn{(\tilde{\Psi}_2,{\cal{N}}(f^u))=p_1(0)B|A|^2+p_2(0)B|B|^2
+\delta_{k_2,2k_1}q(0) A^2+\delta_{k_2,3k_1}q(0) A^3}\label{eq:Bexp}\\
&&\hspace{0.0in}+ \delta_{k_2,-k_1}
\left[P_2(0)AB^2+
(Q_1(0)+P_3(0))A^\ast|B|^2+Q_3(0)(A^\ast)^2 B^\ast
+Q_1(0)A^\ast|A|^2\right]\nonumber
\end{eqnarray}
with coefficients
\begin{eqnarray}
p_1(0)&=&-{i}\left[\frac{<\partial_v\tilde{\psi}_2,\kappa\cdot(h_3-h_5)>}{k_1}
+\frac{<\partial_v\tilde{\psi}_2,\kappa\cdot h_1>}{k_2}
+\frac{\Gamma_3}{k_2-k_1}<\partial_v\tilde{\psi}_2,\kappa\cdot\psi_1>
\right.\nonumber\\
&&\hspace{3.0in}\left.+\frac{\Gamma_5}{k_2+k_1}
<\partial_v\tilde{\psi}_2,\kappa\cdot\psi_1^\ast>\right]\label{eq:p10}\\
p_2(0)&=&-\frac{i}{k_2}\left[<\partial_v\tilde{\psi}_2,\kappa\cdot(h_2-h_6)>
+\frac{\Gamma_6}{2}<\partial_v\tilde{\psi}_2,\kappa\cdot{\psi}_2^\ast>\right],
\label{eq:p20}\\
q(0)&=&\left\{\begin{array}{lc}
-{i}<\partial_v\tilde{\psi}_2,\kappa\cdot{\psi}_1>/{k_1}&
{\rm if}\hspace{0.15in}k_2=2k_1\\
&\\
-i\left[{<\partial_v\tilde{\psi}_2,\kappa\cdot h_4>}/{k_1}
+{\Gamma_4}<\partial_v\tilde{\psi}_2,\kappa\cdot{\psi}_1>/{2k_1}\right]
&{\rm if}\hspace{0.15in}k_2=3k_1.
\end{array}\right.\label{eq:q0}
\end{eqnarray}
In (\ref{eq:Aexp}) and (\ref{eq:Bexp}) we abbreviate the notation of Section
\ref{sec:expand} letting $r_j(0)$ denote $\partial_{\sigma_j}r(0)$, $P_j(0)$
denote $\partial_{\mu_j}P(0)$ and so forth. For an instability with
complex eigenvalues and a reflection
symmetry, the wavenumbers satisfy $k_2=-k_1$,
and the $\Gamma_5$ terms in $r_2(0)$ and $p_1(0)$ are
omitted. In addition, the reflection symmetry (\ref{eq:hopfref}) implies
various identities: $r_1(0)=p_2(0)$, $r_2(0)=p_1(0)$, $h_1(v)=h_2(-v)$,
$h_3(v)=h_3(-v)^\ast$, $h_4(v)=h_6(-v)$, and $h_5(v)=h_5(-v)$, and
these relate the cubic terms in (\ref{eq:Aexp}) and (\ref{eq:Bexp})
as shown.

The coefficients $h_i(v)$ follow from the second-order terms in
(\ref{eq:Sdotwu}). On the left-hand side of (\ref{eq:Sdotwu}), we have
\begin{equation}
\left.\frac{\partial S}{\partial t}\right|_{f^u} =
\left[\frac{\partial H}{\partial A}\dot{A}+\frac{\partial H}{\partial B}\dot{B}
+cc\right]=
\left[\frac{\partial H}{\partial A}\lambda_1{A}
+\frac{\partial H}{\partial B}\lambda_2{B}+cc\right]+{\cal O}(3),
\label{eq:graph1}
\end{equation}
with the partial derivatives evaluated using (\ref{eq:gexp}).
On the right-hand side of (\ref{eq:Sdotwu}), the leading terms are found in
${\cal L}H$ and
\begin{eqnarray}
{\cal{N}}(f^u)&=&\frac{i|A|^2}{k_1}\kappa\cdot\partial_v(\psi_1^\ast-\psi_1)
+\frac{i|B|^2}{k_2}\kappa\cdot\partial_v(\psi_2^\ast-\psi_2)\\
&&+\left\{
\frac{iA^2}{k_1}\kappa\cdot\partial_v\psi_1e^{i2k_1x}
+\frac{iB^2}{k_2}\kappa\cdot\partial_v\psi_2e^{i2k_2x}
+iAB\left(\frac{\kappa\cdot\partial_v\psi_2}{k_1}
+\frac{\kappa\cdot\partial_v\psi_1}{k_2}\right)e^{i(k_1+k_2)x}
\right.\nonumber\\
&&\left.\hspace{0.3in}
+iAB^\ast\left(\frac{\kappa\cdot\partial_v\psi_2^\ast}{k_1}
-\frac{\kappa\cdot\partial_v\psi_1}{k_2}\right)e^{i(k_1-k_2)x}
+cc\right\}
+{\cal O}(3).\nonumber
\end{eqnarray}
The second-order solution of (\ref{eq:Sdotwu}) determines the coefficients in
(\ref{eq:gexp}):
\begin{eqnarray}
h_1(v)&=&\frac{i\kappa\cdot\partial_v(\psi_1^\ast-\psi_1)}{2\gamma_1k_1}
\hspace{0.65in}
h_2(v)=\frac{i\kappa\cdot\partial_v(\psi_2^\ast-\psi_2)}{2\gamma_2k_2}
\label{eq:h1,2}\\
h_3(v)&=&i\,R_{k_2-k_1}(\lambda_1^\ast+\lambda_2)I_3(v)\hspace{0.3in}
h_4(v)=i\,R_{2k_1}(2\lambda_1)I_4(v)\\
h_5(v)&=&i\,R_{k_2+k_1}(\lambda_1+\lambda_2)I_5(v)\hspace{0.3in}
h_6(v)=i\,R_{2k_2}(2\lambda_2)I_6(v)\label{eq:h5,6}
\end{eqnarray}
where $R_{l}(w)$ denotes the resolvent operator (\ref{eq:resol}), and
\begin{eqnarray}
I_3(v)\equiv\frac{\kappa\cdot\partial_v\psi_1^\ast}{k_2}
-\frac{\kappa\cdot\partial_v\psi_2}{k_1}+i\,s(0)\,\psi_1\,\delta_{k_2,2k_1}
&\hspace{0.15in}&
I_5(v)\equiv\frac{\kappa\cdot\partial_v\psi_1}{k_2}
+\frac{\kappa\cdot\partial_v\psi_2}{k_1}\\
I_4(v)\equiv\frac{\kappa\cdot\partial_v\psi_1}{k_1}
+i\,q(0)\,\psi_2\,\delta_{k_2,2k_1}\hspace{0.75in}
&\hspace{0.15in}&
I_6(v)\equiv\frac{\kappa\cdot\partial_v\psi_2}{k_2}.
\end{eqnarray}
These expressions are valid for each of the three instabilities we consider,
with one exception. In the case of complex eigenvalues and
reflection symmetry, when $k_1+k_2=0$, the result for  $h_5(v)$ in
(\ref{eq:h5,6}) is replaced by
\begin{equation}
h_5(v)=-\frac{I_5(v)}{2kz_0}\hspace{0.5in}(k_1+k_2=0)\label{eq:h5}
\end{equation}
in the notation of Section \ref{subsubsec:complex}.

Note that (\ref{eq:h1,2}) implies $\Gamma_1=\Gamma_2=0$ in general, and
(\ref{eq:h5}) forces $\Gamma_5=0$ for this specific type of instability.
Following (I),
we can re-express $(\Gamma_j,h_j)$, $j=3,4, 5, 6$ in (\ref{eq:h1,2}) -
(\ref{eq:h5,6}) more conveniently as
\begin{eqnarray}
\Gamma_j=\left(\frac{1/l_j}{\Lambda_{l_j}(z_j)}\right)\int^\infty_{-\infty}
\frac{dv\sum_s I_j^{(s)}(v)}{v-z_j}&\hspace{0.3in}&
h_j(v)=\frac{I_j(v)}{l_j(v-z_j)}
-\Gamma_j\left(\frac{\kappa\cdot\eta_{l_j}(v)}{v-z_j}\right),\label{eq:g5}
\end{eqnarray}
where
\begin{eqnarray}
l_3=k_2-k_1,\hspace{0.2in}z_3=\frac{k_2z_2-k_1z_1^\ast}{k_2-k_1};
&\hspace{0.1in}&l_4=2k_1,\hspace{0.2in}z_4=z_1;\label{eq:z34}\\
l_5=k_2+k_1,\hspace{0.2in}z_5=\frac{k_2z_2+k_1z_1}{k_2+k_1};
&\hspace{0.1in}&l_6=2k_2,\hspace{0.2in}z_6=z_2.\label{eq:z56}
\end{eqnarray}

\subsection{Previous results for the single mode
instabilities}\label{sec:sing1m}

For $A=0$ in (\ref{eq:adef}) - (\ref{eq:bdef}), we recover the single mode
instability $\dot{B}=p(|B|^2)B=[\lambda_2+p_2(0)|B|^2+\cdots]B$ previously
considered\cite{jdcaj97,jdcaj96}; the asymptotic form of the
cubic coefficient is
\begin{eqnarray}
p_2(0)&=&\frac{1}{\gamma_2^4}
\left[c(k_2,z_2)-\gamma_2\,d(k_2,z_2) + {\cal O}(\gamma_2^2)\right]
\label{eq:p1asymp}
\end{eqnarray}
with the nonsingular functions $c$ and $d$ defined by
\begin{eqnarray}
c(k_2,z_2)&=&-\frac{k_2}{4\Lambda_{k_2}'(z_2)}
{\sum_s}'{\kappa^{(s)}(1-{\kappa^{(s)}}^2)}\;{\rm
Im}\left(\int^\infty_{-\infty}\,dv\frac{\eta_{k_2}^{(s)}}{v-z_2}\right)
\label{eq:a1}\\
d(k_2,z_2)&=&\frac{1}{4}-
\frac{1}{4\Lambda_{k_2}'(z_2)}{\sum_s}'\kappa^{(s)}(1-{\kappa^{(s)}}^2)
\int^\infty_{-\infty}\,dv\frac{\eta_{k_2}^{(s)}}{(v-z_2)^2},\label{eq:b1}
\end{eqnarray}
where the primed species sum omits the electrons.\cite{jdcaj97}
For $n_1>1$ in (\ref{eq:adef}) - (\ref{eq:bdef}), setting $B=0$ determines a
single mode
instability at the longer wavelength:
$\dot{A}=r(|A|^2)A=[\lambda_1+r_1(0)|A|^2+\cdots]A$. The asymptotic form of
$r_1(0)$ follows from (\ref{eq:p1asymp}) - (\ref{eq:b1}) with the replacements
$(\gamma_2,k_2,z_2)\rightarrow(\gamma_1,k_1,z_1)$. We also reduce to
this previously studied case on setting either $A=0$ or $B=0$ in equation
(\ref{eq:o2hopf}); these two limits correspond to single-mode instabilities
in the form of travelling waves propagating in the positive and
negative $x$ directions, respectively.


\section{Singularities in the mode-mode couplings}\label{sec:sing2m}

The central result of the single mode analysis proves that setting
$\beta_j=5/2$ in (\ref{eq:scalings}) yields rescaled amplitude equations
for $a(\tau)$ and $b(\tau)$ that are finite to all orders.\cite{jdcaj97}
When there are two unstable modes additional coupling terms between
the modes are present that have
not been previously considered. In particular in the expansions
(\ref{eq:Aexp}) and (\ref{eq:Bexp}) we find four such couplings,
$(q(0), s(0), p_1(0),r_2(0))$, for
instabilities with $k_2>k_1>0$ (cf. Sections \ref{subsec:complex} -
\ref{subsubsec:real}) and five couplings,
$(p_1(0), Q_1(0), P_2(0), Q_3(0), Q_1(0)+P_3(0))$,
for an instability with $k_2+k_1=0$ (cf. Section \ref{subsubsec:complex}).
The singularities of such mode coupling terms will determine
if the presence of a second unstable wave can alter the nonlinear scaling
associated with single wave instabilities.

This question needs to be formulated carefully to avoid ``trivial''
limits since there are now two distinct linear growth rates $\gamma_1$
and $\gamma_2$. If one growth rate vanishes while the second remains bounded
above zero, then only the singularities associated with the resonant
denominators of the first mode will emerge. This effectively recovers
the singularity structure of the one mode problem even though both
mode amplitudes are non-zero. The more interesting limit arises when all
resonant denominators come into play which requires both growth rates
to vanish simultaneously. Thus we set $\gamma_1=\gamma_2\equiv \gamma$
in the following discussion; in practice this arrangement would
be hard to realize experimentally but could be achieved in a numerical
simulation by simultaneously adjusting the parameters of the equilibrium
and the length of the system.

The origin of the singularities is the same as in the one mode problem:
poles from resonant denominators can straddle the contour of integration
and produce pinching singularities as $\gamma\rightarrow0^+$. Following
the methodology of (I), the worst possible singularity of a given
integral can be estimated by simply counting the total number of poles
(including multiplicity).  This gives an upper bound on the possible
divergence of a given integral which must be checked by a detailed evaluation
once the most divergent integrals have been identified. Ultimately we
are most interested in possible singularities that are stronger than those
already identified in the single-mode subsystem (\ref{eq:p1asymp}).
For example, third order terms with divergences weaker than the $\gamma^{-4}$
singularity of (\ref{eq:p1asymp}) are clearly sub-dominant and cannot alter
the $\beta_j=5/2$ scaling forced by the single mode singularities.

In Sections \ref{sub:universal} - \ref{sub:rescoup}, we assume the single mode
problems exhibit the $\gamma^{-4}$
cubic singularity in (\ref{eq:p1asymp}), i.e.
\begin{equation}
{\sum_s}'{\kappa^{(s)}(1-{\kappa^{(s)}}^2)}\;\eta_{k_j}^{(s)}(v_j)\neq0,
\hspace{0.3in}j=1,2\label{eq:generic}
\end{equation}
for modes $k_1$ and $k_2$, respectively. This is simply the condition that
in (\ref{eq:a1}) $c(k_j,v_j+i0)\neq0$, $j=1,2$. Special limits, such as
infinitely massive ions, for which the single mode system is less singular are
discussed in Section \ref{subsec:special}. Tables 1 and 2 provide a summary
of the asymptotic behavior of the second and third order coupling coefficients.

\subsection{The universal couplings $p_1(0)$ and $r_2(0)$}\label{sub:universal}

The coefficients $r_2(0)$ and  $p_1(0)$ are present at
third order for each instability, and their
asymptotic singularities are fundamental. We discuss the
three types of instabilities separately.

\subsubsection{$F_0(v)\neq F_0(-v)$, complex eigenvalues}\label{subsub:nosym}

For instabilities without reflection symmetry (Section \ref{subsec:complex}),
we have $k_2+k_1>0$ for two positive and unequal wavenumbers.
We first identify integrals in $r_2(0)$ and  $p_1(0)$ with poles above
{\em and} below the contour; terms without this feature are manifestly free
of pinching singularities and will be finite as $\gamma\rightarrow0^+$. In
addition, singularities weaker than $\gamma^{-4}$ are sub-dominant.

{}From (\ref{eq:z34}) - (\ref{eq:z56}), ${\rm Im}(z_j)\geq0$ for $j=1,\ldots,6$
so $\psi_1(v)$, $\psi_2(v)$
have poles only in the upper-half plane as do the conjugated adjoints
$\tilde{\psi}_j(v)^\ast$; thus $<\partial_v\tilde{\psi}_1,\kappa\cdot\psi_2>$
and $<\partial_v\tilde{\psi}_2,\kappa\cdot\psi_1>$ are nonsingular. Similarly,
$I_5(v)$
and $h_5(v)$ have poles only in the
upper-half plane, thus the integral in (\ref{eq:g5}) for $\Gamma_5$ is
nonsingular;
in addition, $\Lambda_{l_5}(z_5)=\Lambda_{k_2+k_1}(z_5)$ is of order unity
as $\gamma\rightarrow0$ since there are no modes at wavenumber $k_2+k_1$ by
assumption.
Hence $\Gamma_5$  and $<\partial_v\tilde{\psi}_1,h_5>$ are both nonsingular.
The integrals $<\partial_v\tilde{\psi}_1,\kappa\cdot\psi_2^\ast>$ and
$<\partial_v\tilde{\psi}_2,\kappa\cdot\psi_1^\ast>$ are nonsingular
unless $v_1=v_2$ in
which case $\gamma^{-2}$ is their worst possible divergence,
so the terms involving $\Gamma_5$ are sub-dominant.

Similarly in
\begin{equation}
<\partial_v\tilde{\psi}_1,\kappa\cdot h_2>=
\frac{i}{2\gamma k_2}\left[
<\partial_v\tilde{\psi}_1,\kappa^2\cdot\partial_v\psi_2^\ast>
-<\partial_v\tilde{\psi}_1,\kappa^2\cdot\partial_v\psi_2>\right]
\label{eq:h2int2}
\end{equation}
the second integral is nonsingular and
the first integral,
$<\partial_v\tilde{\psi}_1,\kappa^2\cdot\partial_v\psi_2^\ast>$,
is nonsingular unless $v_2=v_1$ in which case there is a $\gamma^{-3}$
singularity. This conclusion applies equally to the corresponding terms in
$<\partial_v\tilde{\psi}_2,\kappa\cdot h_1>$; thus these
terms contribute at most a $\gamma^{-4}$ singularity to
$r_2(0)$ and  $p_1(0)$, respectively.

Discarding these terms, we must still consider
the asymptotic behavior due to $h_3$ and $\Gamma_3$:
\begin{eqnarray}
r_2(0)&=&-i\left[\frac{<\partial_v\tilde{\psi}_1,\kappa\cdot h_3^\ast>}{k_2}
-\frac{\Gamma_3^\ast}{k_2-k_1}<\partial_v\tilde{\psi}_1,\kappa\cdot \psi_2>
\right]+\cdots
\label{eq:r2sing}\\
p_1(0)&=&-{i}\left[\frac{<\partial_v\tilde{\psi}_2,\kappa\cdot h_3>}{k_1}
+\frac{\Gamma_3}{k_2-k_1}<\partial_v\tilde{\psi}_2,\kappa\cdot \psi_1>\right]
+\cdots,\label{eq:p1sing}
\end{eqnarray}
where
\begin{eqnarray}
\lefteqn{\Gamma_3=\left(\frac{1/l_3}{\Lambda_{l_3}(z_3)}\right)
\int^\infty_{-\infty}
\frac{dv\sum_s}{v-z_3}
\left[\frac{\kappa^{(s)}\partial_v{\psi_1^{(s)}}^\ast}{k_2}
-\frac{\kappa^{(s)}\partial_v\psi_2^{(s)}}{k_1}
+i\,s(0)\,\psi_1^{(s)}\delta_{k_2,2k_1}\right]}
\label{eq:g3}\hspace{1.0in}\\
<\partial_v\tilde{\psi}_1,\kappa\cdot h_3^\ast>&=&
<\partial_v\tilde{\psi}_1,\frac{\kappa\cdot I_3^\ast}{l_3(v-z_3^\ast)}>
-\Gamma_3^\ast
<\partial_v\tilde{\psi}_1,\frac{\kappa^2\cdot\eta_{l_3}}{(v-z_3^\ast)}>
\label{eq:h3int}\\
<\partial_v\tilde{\psi}_2,\kappa\cdot h_3>&=&
<\partial_v\tilde{\psi}_2,\frac{\kappa\cdot I_3}{l_3(v-z_3)}>
-\Gamma_3<\partial_v\tilde{\psi}_2,\frac{\kappa^2\cdot\eta_{l_3}}{(v-z_3)}>.
\label{eq:h3int2}
\end{eqnarray}
For (\ref{eq:r2sing}) - (\ref{eq:p1sing}), we discuss the
instabilities with $k_2\neq 2k_1$ and $k_2= 2k_1$ separately; the latter
case is more interesting since there can be new singularities when the
modes have the same phase velocity, e.g. in a beam-plasma instability with
a sufficiently cold beam.\cite{jdc95b}

When $k_2\neq 2k_1$, the third term in
(\ref{eq:g3}) is absent and the second term  is manifestly nonsingular;
from (\ref{eq:z34}),
\begin{equation}
z_3=\frac{k_2v_2-k_1v_1}{k_2-k_1}+i\left(\frac{2\gamma}{k_2-k_1}\right),
\end{equation}
so the first term in (\ref{eq:g3})
has a pinching singularity only if $v_2=v_1$ and this possibility yields a
divergence of $\gamma^{-2}$. Also,
$\Lambda_{l_3}(z_3)\equiv\Lambda_{k_2-k_1}(z_3)$ is of order unity
as $\gamma\rightarrow0$ since there are no roots with ${\rm Im}(z)\geq0$
for wavenumbers other than $k_2$ and $k_1$; hence
 the $\Gamma_3$ terms
in (\ref{eq:r2sing}) - (\ref{eq:p1sing}) are sub-dominant. The second
term in (\ref{eq:h3int2}) exhibits only the singularities of $\Gamma_3$ and
may be dropped, while the second term in (\ref{eq:h3int}) has an additional
factor with a pinching singularity similar to $\Gamma_3$, except that the
roles of $z_1$ and $z_3$ are reversed, and gives an overall divergence of
at most $\gamma^{-4}$ (when $v_2=v_1$). The remaining terms in
(\ref{eq:h3int}) - (\ref{eq:h3int2}) are
\begin{eqnarray}
<\partial_v\tilde{\psi}_1,\frac{\kappa\cdot I_3^\ast}{l_3(v-z_3^\ast)}>&=&
<\partial_v\tilde{\psi}_1,
\frac{\kappa^2\cdot\partial_v\psi_1}{k_2l_3(v-z_3^\ast)}>
-<\partial_v\tilde{\psi}_1,
\frac{\kappa^2\cdot\partial_v\psi_2^\ast}{k_1l_3(v-z_3^\ast)}>
\label{eq:I3int}\\
<\partial_v\tilde{\psi}_2,\frac{\kappa\cdot I_3}{l_3(v-z_3)}>&=&
<\partial_v\tilde{\psi}_2,
\frac{\kappa^2\cdot\partial_v\psi_1^\ast}{k_2l_3(v-z_3)}>
-<\partial_v\tilde{\psi}_2,
\frac{\kappa^2\cdot\partial_v\psi_2}{k_1l_3(v-z_3)}>;
\label{eq:I3int2}
\end{eqnarray}
in all cases there are five poles in the integrand and hence a maximum possible
$\gamma^{-4}$ divergence. A closer examination shows that, if $v_2=v_1$,
this singularity is
realized by the first term in (\ref{eq:I3int2}) and both
terms in (\ref{eq:I3int}); in any event
we do not encounter a singularity in $r_2(0)$ or $p_1(0)$ that dominates
those found at third order in the
single mode equations, i.e. a singularity stronger than $\gamma^{-4}$.

When $k_2= 2k_1$,
$\Lambda_{l_3}(z_3)\equiv\Lambda_{k_1}(z_3)$ with
\begin{equation}
z_3=({2v_2-v_1})+i\left(\frac{2\gamma}{k_1}\right).
\end{equation}
Thus when $v_2=v_1$ we have $z_3=z_1+i\gamma/k_1$, and $\Lambda_{l_3}(z_3)$
is ${\cal O}(\gamma)$ as $\gamma\rightarrow0^+$; if $v_2\neq v_1$,
$\Lambda_{l_3}(z_3)$ is still of order one. Thus with equal phase velocities,
the singularity of $\Gamma_3$ is increased to $\gamma^{-3}$
(including the presence of the third term in (\ref{eq:g3}) which
has a $\gamma^{-2}$ divergence in $s(0)$); nevertheless
the $\Gamma_3$ terms in (\ref{eq:r2sing}) - (\ref{eq:p1sing}) are still
sub-dominant. The second term in (\ref{eq:h3int2}) can be neglected for similar
reasons. However,
the second term in (\ref{eq:h3int}) now has an apparent divergence
of $\gamma^{-5}$ when $v_2=v_1$; such a singularity would
not be absorbed by the scalings
used to remove the $\gamma^{-4}$ singularities characteristic of the single
mode problem. We will evaluate this term more precisely shortly.
The first terms in (\ref{eq:h3int}) - (\ref{eq:h3int2}) contain extra
contributions from the third term in (\ref{eq:g3}):
\begin{eqnarray}
<\partial_v\tilde{\psi}_1,\frac{\kappa\cdot I_3^\ast}{l_3(v-z_3^\ast)}>&=&
<\partial_v\tilde{\psi}_1,
\frac{\kappa^2\cdot\partial_v\psi_1}{k_2l_3(v-z_3^\ast)}>
-<\partial_v\tilde{\psi}_1,
\frac{\kappa^2\cdot\partial_v\psi_2^\ast}{k_1l_3(v-z_3^\ast)}>
\label{eq:I3intr}\\
&&\hspace{1.0in}-i\,s(0)^\ast
<\partial_v\tilde{\psi}_1,
\frac{\kappa\cdot \psi_1^\ast}{l_3(v-z_3^\ast)}>\nonumber\\
<\partial_v\tilde{\psi}_2,\frac{\kappa\cdot I_3}{l_3(v-z_3)}>&=&
<\partial_v\tilde{\psi}_2,
\frac{\kappa^2\cdot\partial_v\psi_1^\ast}{k_2l_3(v-z_3)}>
-<\partial_v\tilde{\psi}_2,
\frac{\kappa^2\cdot\partial_v\psi_2}{k_1l_3(v-z_3)}>
\label{eq:I3intr2}\\
&&\hspace{1.0in}+i\,s(0)
<\partial_v\tilde{\psi}_2,
\frac{\kappa\cdot \psi_1}{l_3(v-z_3)}>.\nonumber
\end{eqnarray}
Our previous discussion of the first two terms in (\ref{eq:I3intr})
and (\ref{eq:I3intr2}) is unchanged, and the new term in (\ref{eq:I3intr2})
shows only the $\gamma^{-2}$ singularity in $s(0)$ (as determined from
equation (\ref{eq:s0asym}) below). This singularity is
also present in the third term of (\ref{eq:I3intr}),
but now $s(0)^\ast$ multiplies
an integral that has an apparent singularity of $\gamma^{-3}$ giving
a second term with overall $\gamma^{-5}$ divergence.

The foregoing discussion shows that there are new singularities
in the coupling coefficients and that these singularities are most severe
when $k_2=2k_1$ and $v_1=v_2$. In this resonant case, the $p_1(0)$
singularity never exceeds $\gamma^{-4}$, but
we have identified two
contributions to $<\partial_v\tilde{\psi}_1,h_3^\ast>$ in the
cubic coefficient $r_2(0)$ whose singularities
appear to dominate the $\gamma^{-4}$ divergence
characteristic of the cubic terms of the single mode problem.
We proceed to
a detailed evaluation of these exceptional terms which shows that the
$\gamma^{-5}$ singularities are typically present.

For the second term in (\ref{eq:h3int}) the calculation of the integrals
using partial fraction expansions (cf. (I)) yields the following asymptotic
forms when $v_2=v_1$:
\begin{eqnarray}
\Gamma_3&=&\frac{1}{\gamma^3}
\left[\left(\frac{2\pi k_1^2}{9\Lambda_{k_1}'(v_1)}\right)
{\sum_s}'\kappa^{(s)}(1+\kappa^{(s)})
\eta_{k_1}^{(s)}(v_1)+{\cal O}(\gamma)\right]\label{eq:newsinga}\\
<\partial_v\tilde{\psi}_1,\frac{\kappa^2\cdot\eta_{l_3}}{(v-z_3^\ast)}>&=&
\frac{1}{\gamma^2}
\left[\left(\frac{2\pi i\,k_1^2}{9\Lambda_{k_1}'(v_1)}\right)
{\sum_s}'\kappa^{(s)}(1+\kappa^{(s)}) \eta_{k_1}^{(s)}(v_1)
+{\cal O}(\gamma)\right],\label{eq:newsingb}
\end{eqnarray}
where the primed species summation excludes the electrons. In writing
(\ref{eq:newsingb}) we have used the dispersion relation (\ref{eq:critroot}).
Given the assumption (\ref{eq:generic}) on the single mode problem, we expect
${\sum_s}'\kappa^{(s)}(1+\kappa^{(s)}) \eta_{k_1}^{(s)}(v_1)\neq0$
to typically hold; hence the $\gamma^{-5}$ singularity is generally present.
Finally, the third term in (\ref{eq:I3intr}) contains a $\gamma^{-2}$
singularity from $s(0)^\ast$ (equation (\ref{eq:s0asym})) while a partial
fraction
expansion of the  integral yields
\begin{eqnarray}
<\partial_v\tilde{\psi}_1,\frac{\kappa\cdot\psi_1^\ast}{(v-z_3^\ast)}>&=&
\frac{1}{\gamma^3}\left[-
\left(\frac{5\pi k_1^3} {18\Lambda_{k_1}'(v_1)}\right)
{\sum_s}'\kappa^{(s)}(1+\kappa^{(s)}) \eta_{k_1}^{(s)}(v_1)
+{\cal O}(\gamma)\right].\label{eq:newsingc}
\end{eqnarray}
Thus the third term in (\ref{eq:I3intr}) also realizes a $\gamma^{-5}$
singularity. These singularities require a shift in the scaling exponents
that characterize the single mode instability; this point is discussed below in
Section \ref{sec:nonlinear}.

\subsubsection{$F_0(v)= F_0(-v)$, real eigenvalues}

For reflection-symmetric instabilities with real eigenvalues (Section
\ref{subsubsec:real}),
we have $v_1=v_2=0$ and $k_2+k_1>0$ for two positive and unequal wavenumbers.
The previous analysis is applicable here and we obtain the same conclusions
with one modification. Since the condition $v_1=v_2$ is automatically
satisfied, the ``spatial resonance'', $k_2=2k_1$, is sufficient to obtain
the extra singularities noted above. These results are summarized in Table 1.
\begin{table}
\caption{Generic singularities of the couplings and scaling exponents}
\vspace{0.25in}
\begin{tabular}{lcccccccc}
Instability & Resonance&$q$ &$s$&$r_1^\dagger$&$r_2$&$p_1$&
$p_2^\dagger$&$(\beta_1,\beta_2)$\\
\hline\\
$F_0(v)\neq F_0(-v)$&$k_2\neq2k_1$, $v_2\neq v_1$&- & - &$\gamma^{-4}$ &
$\gamma^{-1}$&$\gamma^{-1}$&$\gamma^{-4}$&
$(\frac{5}{2},\frac{5}{2})$\vspace{0.05in}\\
$k_2>k_1>0$ &\hspace{0.6in} $v_2= v_1$&- &-& $\gamma^{-4}$&
$\gamma^{-4}$&$\gamma^{-4}$&$\gamma^{-4}$&
$(\frac{5}{2},\frac{5}{2})$\\
$\lambda_1,\lambda_2$ complex&&&   & &&&&\\
&$k_2=2k_1$, $v_2\neq v_1$&
${\cal O}(1)$  &$\gamma^{-2}$&$\gamma^{-4}$ &$\gamma^{-2}$&
$\gamma^{-2}$&$\gamma^{-4}$&$(\frac{5}{2},\frac{5}{2})$
\vspace{0.05in}\\
&\hspace{0.6in} $v_2= v_1$&${\cal O}(1)$&$\gamma^{-2}$& $\gamma^{-4}$&
$\gamma^{-5}$& $\gamma^{-4}$&$\gamma^{-4}$&
$(\frac{5}{2},3)$\vspace{0.05in}\\
&&&   & &&&&\\
&$k_2=3k_1$, $v_2\neq v_1$&
${\cal O}(1)$  &$\gamma^{-4}$&$\gamma^{-4}$ &$\gamma^{-1}$&
$\gamma^{-1}$&$\gamma^{-4}$&$(\frac{5}{2},\frac{5}{2})$
\vspace{0.05in}\\
&\hspace{0.6in} $v_2= v_1$&${\cal O}(1)$&$\gamma^{-4}$& $\gamma^{-4}$&
$\gamma^{-4}$& $\gamma^{-4}$&$\gamma^{-4}$&
$(\frac{5}{2},\frac{5}{2})$\vspace{0.05in}\\
\hline\\
$F_0(v)= F_0(-v)$&$k_2\neq2k_1$&-& - &$\gamma^{-4}$ &
$\gamma^{-4}$&$\gamma^{-4}$&$\gamma^{-4}$&
$(\frac{5}{2},\frac{5}{2})$\\
$k_2>k_1>0$ &&&& && &&\\
$\lambda_1,\lambda_2$ real,&$k_2=2k_1$&${\cal O}(1)$&$\gamma^{-2}$&
$\gamma^{-4}$ &$\gamma^{-5}$&
$\gamma^{-4}$&$\gamma^{-4}$&
$(\frac{5}{2},3)$\\
multiplicity-two&&&&&&&&\\
&$k_2=3k_1$&${\cal O}(1)$&$\gamma^{-4}$&
$\gamma^{-4}$ &$\gamma^{-4}$&
$\gamma^{-4}$&$\gamma^{-4}$&
$(\frac{5}{2},\frac{5}{2})$\\
\hline\\
$\dagger=$ single mode &  & &&&&\\
\end{tabular}
\label{table1}
\end{table}

\subsubsection{$F_0(v)= F_0(-v)$, complex eigenvalues}

For reflection-symmetric instabilities with complex eigenvalues (Section
\ref{subsubsec:complex}), we have $v_1=-v_2$ and $k_2+k_1=0$. These
conditions rule out the presence of $k_2=2k_1$ or $k_2=3k_1$
resonances, and imply that $p_1(0)=r_2(0)$, where
\begin{equation}
p_1(0)=-{i}\left[\frac{<\partial_v\tilde{\psi}_2,\kappa\cdot (h_3-h_5)>}{k_1}
+\frac{<\partial_v\tilde{\psi}_2,\kappa\cdot h_1>}{k_2}
+\frac{\Gamma_3<\partial_v\tilde{\psi}_2,\kappa\cdot \psi_1>}{k_2-k_1}
\right]
\end{equation}
with $h_5$ defined in (\ref{eq:h5}).

{}From $z_1=-z_2=z_0$ we now have $z_1$ in the upper-half plane ($k_1>0$) and
$z_2$ in the lower-half plane ($k_2<0$) along with $z_3=-i\gamma/k_1$.
The eigenfunctions $\psi_1(v)$ and $\tilde{\psi}_1(v)^\ast$ have poles in the
upper-half plane that approach $v_1$ as $\gamma\rightarrow0^+$, while the poles
of $\psi_2(v)$ and $\tilde{\psi}_2(v)^\ast$ are in the lower half-plane and
approach $v_2$ as $\gamma\rightarrow0^+$. The integrals
$<\partial_v\tilde{\psi}_1,\kappa\cdot \psi_2>$
and $<\partial_v\tilde{\psi}_2,\kappa\cdot \psi_1>$
are still nonsingular although
they now involve poles above and below the contour. The relation $v_1=-v_2$
ensures that no pinching singularity develops.
Similarly, $I_5(v)$ and $h_5(v)$ have poles on each side of the contour
but no pinch can develop in the integrals
$<\partial_v\tilde{\psi}_1,\kappa\cdot h_5>$
and $<\partial_v\tilde{\psi}_2,\kappa\cdot h_5>$. In
\begin{equation}
<\partial_v\tilde{\psi}_1,\kappa\cdot h_2>=
\frac{i}{2\gamma k_2}\left[
<\partial_v\tilde{\psi}_1,\kappa^2\cdot\partial_v\psi_2^\ast>
-<\partial_v\tilde{\psi}_1,\kappa^2\cdot\partial_v\psi_2>\right],
\end{equation}
the first integral
$<\partial_v\tilde{\psi}_1,\kappa^2\cdot\partial_v\psi_2^\ast>$
only has poles in the upper-half plane while the second integral has poles
above
and below the contour but no pinch; thus this term may also be dropped as well
as the corresponding
term $<\partial_v\tilde{\psi}_2,\kappa\cdot h_1>$ in $p_1(0)$.

Discarding these manifestly nonsingular terms, we must still reconsider
the remaining terms in (\ref{eq:r2sing}) - (\ref{eq:h3int2}). The prior
analysis given in Section \ref{subsub:nosym} requires relatively minor
modifications to allow for the relation $v_1=-v_2$ and the shifted
location of the poles $z_2$ and $z_3$. In (\ref{eq:g3}),
$\Lambda_{l_3}(z_3)\equiv\Lambda_{-2k_1}(z_3)$ is of order one as
$\gamma\rightarrow0^+$
and the integrals are free of pinching singularities since $z_3$ is pure
imaginary and
can never converge to either $z_1$ or $z_2$. Thus the second terms in
(\ref{eq:r2sing}) - (\ref{eq:p1sing}) are nonsingular as are
the second terms in (\ref{eq:h3int}) - (\ref{eq:h3int2}). The final integrals
in the first terms of
(\ref{eq:h3int}) - (\ref{eq:h3int2}), written out in (\ref{eq:I3int}) -
(\ref{eq:I3int2}), are also manifestly free of pinching
singularities. In summary, the coupling coefficients $p_1(0)$, $r_2(0)$
for instabilities with complex coefficients and reflection-symmetry
are identical, and exhibit singularities due to the explicit
$\gamma^{-1}$ factor in $<\partial_v\tilde{\psi}_1,\kappa\cdot h_2>$ only.
\begin{table}
\caption{Generic singularities of the couplings and scaling exponents}
\vspace{0.25in}
\begin{tabular}{lccccccccc}
Instability & $Q_1$&$P_2$ &$Q_3$&$(Q_1+P_3)$&$r_1^\dagger$&$r_2$&$p_1$&
$p_2^\dagger$&$(\beta_1,\beta_2)$\\
\hline\\
$F_0(v)= F_0(-v)$&$\gamma^{-1}$&${\cal O}(1)$&
$\gamma^{-3}$  &$\gamma^{-4}$&$\gamma^{-4}$ &
$\gamma^{-1}$&$\gamma^{-1}$&$\gamma^{-4}$&
$(\frac{5}{2},\frac{5}{2})$\\
$k_2+k_1=0$ &&&& && &&\\
$\lambda_1$ complex,&&& &  & &&&&\\
multiplicity-two&&& &&&&&&\\
\hline\\
$\dagger=$ single mode &  & &&&&&\\
\end{tabular}
\label{table2}
\end{table}

\subsection{The couplings $Q_1(0)$, $P_2(0)$, $Q_3(0)$ and $Q_1(0)+P_3(0)$
}\label{sub:zero}

For  reflection-symmetric instabilities with complex eigenvalues (Section
\ref{subsubsec:complex}), there are four additional O(2) symmetric
couplings at third order. Although these terms can be removed by a coordinate
transformation to obtain the normal form in (\ref{eq:o2hopf}),
it is important to consider their asymptotic behavior.

The singularities for the couplings in this instablity are summarized in
Table 2. $P_2(0)$ is nonsingular and the singularities of $Q_1(0)$ and
$Q_3(0)$ are sub-dominant. The strongest singularity occurs in $Q_1(0)+P_3(0)$
due the integral
\begin{equation}
<\partial_v\tilde{\psi}_1,\kappa\cdot h_1>=
\frac{i}{2\gamma k_1}\left[
<\partial_v\tilde{\psi}_1,\kappa^2\cdot\partial_v\psi_1^\ast>
-<\partial_v\tilde{\psi}_1,\kappa^2\cdot\partial_v\psi_1>\right],
\end{equation}
whose first term
$<\partial_v\tilde{\psi}_1,\kappa^2\cdot\partial_v\psi_1^\ast>$
has a $\gamma^{-3}$ singularity giving an overall singularity of
$\gamma^{-4}$ for $Q_1(0)+P_3(0)$.

\subsection{The spatial resonances: $q(0)$ and $s(0)$}\label{sub:rescoup}

When $k_2=2k_1$ and $k_2=3k_1$ there are additional couplings   at
second and third order, respectively. We first consider $q(0)$ and $s(0)$
for the $k_2=2k_1$
resonance, noting that the two integrals,
$<\partial_v\tilde{\psi}_1,\kappa\cdot\psi_2>$ and
$<\partial_v\tilde{\psi}_2,\kappa\cdot\psi_1>$, have poles at $z_1$ and $z_2$
and are free of pinching singularities in all cases. Thus $q(0)$ is nonsingular
as is the second term in $s(0)$. The first term in $s(0)$
was evaluated in (I),
\begin{eqnarray}
<\partial_v\tilde{\psi}_1,\kappa\cdot\psi_1^\ast>&=&
\left(\frac{ik_1}{2\gamma_1}\right)^2
\left[\frac{2i}{\Lambda'_{k_1}(z_1)}
{\sum_s}'\kappa^{(s)}(1+\kappa^{(s)})\;{\rm Im}
\int^\infty_{-\infty}\frac{dv\,\eta_{k_1}^{(s)}}
{v-z_1}+{\cal O}(\gamma_1)\right],\label{eq:close}
\end{eqnarray}
where the primed species sum omits the electrons;
this determines the singularity of $s(0)$,
\begin{equation}
s(0)=\frac{1}{\gamma^2}
\left[-\frac{\pi k_1^2}{2\Lambda'_{k_1}(z_1)}
{\sum_s}'\kappa^{(s)}(1+\kappa^{(s)})\;\eta_{k_1}^{(s)}(v_1)
+{\cal O}(\gamma)\right].\label{eq:s0asym}
\end{equation}

For the $k_2=3k_1$ resonance $q(0)$ is again readily seen to be nonsingular
and we omit the details. For $s(0)$ in (\ref{eq:s0}) the
integral $<\partial_v\tilde{\psi}_1,\kappa\cdot h_4^\ast>$ contains
a contribution,
\begin{equation}
<\partial_v\tilde{\psi}_1,\frac{\kappa\cdot I_4^\ast}{v-z_1^\ast}>\,
\sim\, {\cal O}(\gamma^{-4})
\end{equation}
if $v_2\neq v_1$. When $v_2= v_1$ there are additional
singularities, but none stronger than $\gamma^{-4}$. These conclusions are
summarized in Table 1.

\subsection{Special limits: coupling singularities with fixed
ions}\label{subsec:special}

In the various explicit asymptotic formulas, such as (\ref{eq:p1asymp}),
(\ref{eq:newsinga}) - (\ref{eq:newsingc}), and (\ref{eq:s0asym}),
the leading term vanishes if
the ion masses are treated as infinite since $\kappa^{(s)}\rightarrow 0$
in this limit. In (I), this suppression of the leading singularity was shown to
be a general feature of the integrals that appearing in the amplitude
equations.
For the single mode instability, the cubic coefficient $p_1(0)$ has a
$\gamma^{-3}$ singularity when the ions are fixed, and  the modified
single-mode scaling $A(t)=\gamma^{2}a(\gamma t)$
suffices to render the amplitude expansion finite.\cite{jdcaj97,jdc95}

It is straightforward to adapt the results of the previous sections to
the case of infinitely massive ions: with only a few exceptions
among terms that are already sub-dominant the generic divergence is reduced
by one factor of $\gamma^{-1}$. For our purposes it is suffices to summarize
in Table 3 the resulting changes to Table 1 when the
electrons are the only mobile species.
\begin{table}[t]
\caption{Summary of the singularities of the couplings for fixed ions}
\vspace{0.25in}
\begin{tabular}{lcccccccc}
Instability & Resonance&$q$ &$s$&$r_1^\dagger$&$r_2$&$p_1$&
$p_2^\dagger$&$(\beta_1,\beta_2)$\\
\hline\\
$F_0(v)\neq F_0(-v)$&$k_2\neq2k_1$, $v_2\neq v_1$&- & - &$\gamma^{-3}$ &
$\gamma^{-1}$&$\gamma^{-1}$&$\gamma^{-3}$&
$({2},{2})$\vspace{0.05in}\\
$k_2>k_1>0$ &\hspace{0.6in} $v_2= v_1$&- &-& $\gamma^{-3}$&
$\gamma^{-3}$&$\gamma^{-3}$&$\gamma^{-3}$&
$({2},{2})$\\
$\lambda_1,\lambda_2$ complex&&&   & &&&&\\
&$k_2=2k_1$, $v_2\neq v_1$&
${\cal O}(1)$  &$\gamma^{-1}$&$\gamma^{-3}$ &$\gamma^{-1}$&
$\gamma^{-1}$&$\gamma^{-3}$&$({2},{2})$
\vspace{0.05in}\\
&\hspace{0.6in} $v_2= v_1$&${\cal O}(1)$&$\gamma^{-1}$& $\gamma^{-3}$&
$\gamma^{-4}$& $\gamma^{-3}$&$\gamma^{-3}$&
$({2},\frac{5}{2})$\vspace{0.05in}\\
&&&   & &&&&\\
&$k_2=3k_1$, $v_2\neq v_1$&
${\cal O}(1)$  &$\gamma^{-3}$&$\gamma^{-3}$ &$\gamma^{-1}$&
$\gamma^{-1}$&$\gamma^{-3}$&$({2},{2})$
\vspace{0.05in}\\
&\hspace{0.6in} $v_2= v_1$&${\cal O}(1)$&$\gamma^{-3}$& $\gamma^{-3}$&
$\gamma^{-3}$& $\gamma^{-3}$&$\gamma^{-3}$&
$({2},{2})$\vspace{0.05in}\\
\hline\\
$F_0(v)= F_0(-v)$&$k_2\neq2k_1$&-& - &$\gamma^{-3}$ &
$\gamma^{-3}$&$\gamma^{-3}$&$\gamma^{-3}$&
$({2},{2})$\\
$k_2>k_1>0$ &&&& && &&\\
$\lambda_1,\lambda_2$ real,&$k_2=2k_1$&${\cal O}(1)$&$\gamma^{-1}$&
$\gamma^{-3}$ &$\gamma^{-4}$&
$\gamma^{-3}$&$\gamma^{-3}$&
$({2},\frac{5}{2})$\\
multiplicity-two&&&&&&&&\\
&$k_2=3k_1$&${\cal O}(1)$&$\gamma^{-3}$&
$\gamma^{-3}$ &$\gamma^{-3}$&
$\gamma^{-3}$&$\gamma^{-3}$&
$({2},{2})$\\
\hline\\
$\dagger=$ single mode &  & &&&&\\
\end{tabular}
\label{table3}
\end{table}

\section{Nonlinear scalings}\label{sec:nonlinear}

In this section we make use of the leading order behavior of the coupling
coefficients identified in the preceding section to determine the scaling of
the saturation amplitude of the two competing modes with the growth rate
$\gamma$. We first consider this question for the coefficient singularities
listed in Table 1, i.e. for instabilities with $k_2>k_1>0$.
Following (I), we introduce scaled amplitudes
$A(t)\equiv\gamma^{\beta_1}a(\gamma t)\exp(-i\theta_1(t))$ and
$B(t)\equiv\gamma^{\beta_2}a(\gamma t)\exp(-i\theta_2(t))$ for $\gamma>0$ and
rewrite (\ref{eq:aeqn}) - (\ref{eq:beqn}) using the expansions in
(\ref{eq:Aexp}) and (\ref{eq:Bexp}):
\begin{eqnarray}
\frac{da}{d\tau}&=&a+\gamma^{\beta_2-1}\,{\rm
Re}[s(0)\,e^{i(2\theta_1-\theta_2)}]a\,b\,\delta_{k_2,2k_1}
+\gamma^{2\beta_1-1}\,{\rm Re}[r_1(0)]a^3\label{eq:ascaled}\\
&&\hspace{0.75in}
+\gamma^{2\beta_2-1}\,{\rm Re}[r_2(0)]a\,b^2
+\gamma^{\beta_2+\beta_1-1}\,{\rm
Re}[s(0)\,e^{i(3\theta_1-\theta_2)}]a^2\,b\,\delta_{k_2,3k_1}
+\cdots\nonumber\\
\frac{db}{d\tau}&=&b+\gamma^{2\beta_1-\beta_2-1}\,{\rm
Re}[q(0)\,e^{-i(2\theta_1-\theta_2)}]a^2\,\delta_{k_2,2k_1}
+\gamma^{2\beta_1-1}\,{\rm Re}[p_1(0)]a^2\,b
\label{eq:bscaled}\\
&&\hspace{0.75in}
+\gamma^{3\beta_1-\beta_2-1}\,{\rm
Re}[q(0)\,e^{-i(3\theta_1-\theta_2)}]a^3\,\delta_{k_2,3k_1}
+\gamma^{2\beta_2-1}\,{\rm Re}[p_2(0)]\,b^3+\cdots\nonumber\\
\frac{d\theta_1}{dt}&=&\omega_1-\gamma^{\beta_2}
\,{\rm Im}[s(0)\,e^{i(2\theta_1-\theta_2)}]\,b\,\delta_{k_2,2k_1}
-\gamma^{2\beta_1}\,{\rm Im}[r_1(0)]a^2\label{eq:tascaled}\\
&&\hspace{0.75in}
-\gamma^{2\beta_2}\,{\rm Im}[r_2(0)]\,b^2
-\gamma^{\beta_2+\beta_1}\,{\rm Im}[s(0)\,e^{i(3\theta_1-\theta_2)}]
\,ab\,\delta_{k_2,3k_1}+\cdots\nonumber\\
\frac{d\theta_2}{dt}&=&\omega_2-\gamma^{2\beta_1-\beta_2}
\,{\rm Im}[q(0)\,e^{-i(2\theta_1-\theta_2)}]\frac{a^2}{b}\,\delta_{k_2,2k_1}
-\gamma^{2\beta_1}\,{\rm Im}[p_1(0)]a^2\label{eq:tbscaled}\\
&&\hspace{0.75in}
-\gamma^{2\beta_2}\,{\rm Im}[p_2(0)]\,b^2
-\gamma^{3\beta_1-\beta_2}
\,{\rm Im}[q(0)\,e^{-i(3\theta_1-\theta_2)}]\frac{a^3}{b}\,\delta_{k_2,3k_1}
+\cdots.\nonumber
\end{eqnarray}
If possible, the choice of $\beta_1$ and $\beta_2$ should be made so that
each term is finite as $\gamma\rightarrow0^+$ and there is
a formal balance between linear and nonlinear terms in (\ref{eq:ascaled}) -
(\ref{eq:bscaled}).

In Table 1, we focus initially on the instabilities with $k_2\neq 2k_1$. The
$\gamma^{-4}$ singularities of the single mode
coefficients $r_1(0)$ and $p_2(0)$ in (\ref{eq:ascaled}) - (\ref{eq:bscaled})
require $\beta_1\geq5/2$ and $\beta_2\geq5/2$. These exponents are large enough
to ensure a finite limit for each term (\ref{eq:tascaled}) -
(\ref{eq:tbscaled}); in fact the phase equations
reduce to $\theta_j=\omega_j+{\cal O}(\gamma)$, $j=1,2$. In (\ref{eq:ascaled})
- (\ref{eq:bscaled}), the minimal choice $\beta_j=5/2$ suffices to control the
singularities in the mode couplings $r_2(0)$ and $p_1(0)$ as well
and the amplitude equations reduce to
\begin{eqnarray}
\frac{da}{d\tau}&=&a+{\rm Re}[c(k_1,v_1+i0)]a^3+\gamma^{4}\,{\rm
Re}[r_2(0)]a\,b^2
+\gamma^{4}\,{\rm Re}[s(0)\,e^{i(3\theta_1-\theta_2)}]a^2\,b\,\delta_{k_2,3k_1}
+\cdots\label{eq:ascalednr}\\
\frac{db}{d\tau}&=&b+\gamma^{4}\,{\rm Re}[p_1(0)]a^2\,b+
{\rm Re}[c(k_2,v_2+i0)]\,b^3
+\gamma^{4}\,{\rm Re}[q(0)\,e^{-i(3\theta_1-\theta_2)}]a^3\,\delta_{k_2,3k_1}
+\cdots,
\label{eq:bscalednr}
\end{eqnarray}
where $c(k_j,z_j)$ is defined in (\ref{eq:a1}). In these variables
the linear growth rates are unity and the single mode terms,
$a^3$ and $b^3$, are of
${\cal O}(1)$ in $\gamma$.  If $v_2=v_1$, then the coupling coefficients
$\gamma^4r_2(0)$
and $\gamma^4p_1(0)$ are also of ${\cal O}(1)$, but otherwise these terms are
sub-dominant. When it is present, the resonant term,
$\gamma^4s(0)\,e^{i(3\theta_1-\theta_2)}$, is formally of ${\cal O}(1)$;
however the phases in the exponential
evolve on the fast time scale $t=\tau/\gamma$ and therefore the exponential
oscillates very rapidly as $\gamma\rightarrow0^+$ unless the linear frequencies
are resonant: $3\omega_1=\omega_2$. Such a rapid oscillation allows the term
to be time-averaged and neglected, but when $3\omega_1=\omega_2$ the phase
is stationary and this ${\cal O}(1)$ term cannot be discarded by
time-averaging. Note that, given the spatial resonance $k_2=3k_1$, the
frequency resonance is equivalent to assuming that the linear phase velocities
are equal $v_2=v_1$. The second resonant term,
$\gamma^4q(0)\,e^{-i(3\theta_1-\theta_2)}$, is formally of ${\cal O}(\gamma^4)$
but nevertheless has a qualitatively important effect on the dynamics near
$b=0$ as discussed below. Thus the asymptotic equations (\ref{eq:ascalednr}) -
(\ref{eq:bscalednr}) are correct when
$v_2=v_1$, but if $v_2\neq v_1$ then their form simplifies to
\begin{eqnarray}
\frac{da}{d\tau}&=&a+{\rm Re}[c(k_1,v_1+i0)]a^3+\cdots\label{eq:ascalednr2}\\
\frac{db}{d\tau}&=&b+{\rm Re}[c(k_2,v_2+i0)]\,b^3
+\gamma^{4}\,{\rm
Re}[q(0)\,e^{-i(3\theta_1-\theta_2)}]a^3\,\delta_{k_2,3k_1}+\cdots;
\label{eq:bscalednr2}
\end{eqnarray}
this system describes the evolution of the waves as if they were
non-interacting, indeed their mutual dynamics is essentially the
``superposition'' of the two single mode amplitude equations save
for the very weak resonant
term in (\ref{eq:bscalednr2}).
For $k_2\neq 2 k_1$, our principal conclusion is
that no change from the
scaling exponents predicted by the single mode singularities is indicated.

The resonance $k_2=2k_1$ is special: the $s(0)$ resonant term
now occurs at second
order with a $\gamma^{-2}$ singularity and, if $v_2=v_1$, the singularity in
$r_2(0)$ jumps to $\gamma^{-5}$. In addition,  $v_2=v_1$ implies
a frequency resonance $2\omega_1=\omega_2$ that suppresses the oscillation of
the quadratic term. If we keep $\beta_2=5/2$ in
(\ref{eq:ascaled}), then this quadratic term retains a $\gamma^{-1/2}$
singularity
that cannot be erased by time averaging. Similarly the cubic term
$\gamma^{4}\,{\rm Re}[r_2(0)]a\,b^2$ would retain a $\gamma^{-1}$ singularity.
In this situation, the scaling exponent for the
short wavelength mode must
be increased to $\beta_2=3$ to obtain a sensible asymptotic limit.
Now the amplitude equations reduce to
\begin{eqnarray}
\frac{da}{d\tau}&=&a+\gamma^{2}\,{\rm Re}[s(0)\,e^{i(2\theta_1-\theta_2)}]
a\,b+{\rm Re}[c(k_1,v_1+i0)]a^3+\gamma^{5}\,{\rm
Re}[r_2(0)]a\,b^2+\cdots\label{eq:ascaledr}\\
\frac{db}{d\tau}&=&b+\gamma\,{\rm
Re}[q(0)\,e^{-i(2\theta_1-\theta_2)}]a^2+\gamma^{4}\,{\rm
Re}[p_1(0)]a^2\,b+\cdots;
\label{eq:bscaledr}
\end{eqnarray}
in these variables the growth rates are again unity and all terms shown are
${\cal O}(1)$ in $\gamma$ except for $\gamma\,{\rm
Re}[q(0)\exp{-i(2\theta_1-\theta_2)}]a^2$ which is ${\cal O}(\gamma)$
but is nevertheless important for the asymptotic dynamics as explained below.
When $k_2=2k_1$ but $v_2\neq v_1$, the quadratic term in
(\ref{eq:ascaledr}) can be removed by
time-averaging and the singularity of $r_2(0)$ and $p_1(0)$ drops to
$\gamma^{-2}$; in this
case the single mode scalings suffice to control the singularities in
(\ref{eq:ascaled}) - (\ref{eq:bscaled}) and the amplitude equations simplify
to
\begin{eqnarray}
\frac{da}{d\tau}&=&a+{\rm Re}[c(k_1,v_1+i0)]a^3+\cdots\label{eq:ascalednr3}\\
\frac{db}{d\tau}&=&b+\gamma\,{\rm Re}[q(0)\,e^{-i(2\theta_1-\theta_2)}]a^2
+{\rm Re}[c(k_2,v_2+i0)]\,b^3+\cdots.
\label{eq:bscalednr3}
\end{eqnarray}

In the resonant regime, $k_2=2k_1$ and $v_2=v_1$, the long wavelength mode
saturates at a ${\cal O}(\gamma^{5/2})$ amplitude while the short
wavelength mode
saturates at a much smaller ${\cal O}(\gamma^3)$ amplitude, i.e. the
presence of the long mode results in a dramatic suppression of the amplitude
of the short mode. Because of the $\beta_2=3$ scaling required for the short
wave
mode the equation for $b(\tau)$ is dominated by the interaction with the long
wave
mode $a(\tau)$. In particular in the rescaled equation (\ref{eq:bscaledr}) the
single mode terms
appear at higher order and hence are omitted.
On the other hand we have retained the ${\cal O}(\gamma)$ term $a^2$. Although
formally small this term exerts a profound effect on the resulting dynamics.
This is because it destroys the invariance of the subspace $b=0$. This term
therefore plays the role of a ``noise'' term that continuously forces the
short mode. This observation holds equally for the resonant $a^3$
term in (\ref{eq:bscalednr}).

The singularities summarized in Table 2 refer to an instability of a
reflection-symmetric equilibrium involving complex eigenvalues. All the
couplings in Table 2 are third order terms and none of the
divergences are stronger than the $\gamma^{-4}$ singularities of the
single mode couplings. Thus the single mode exponents $\beta_j=5/2$ will
control all singularities in the amplitude equations to third order, and the
rescaled equations have the form
\begin{eqnarray}
\frac{da}{d\tau}&=&a+{\rm Re}[c(k_1,v_1+i0)]a^3
+\gamma^{4}\,{\rm Re}[e^{i(\theta_b+\theta_a)}(Q_1(0)+P_3(0))]
b a^2+\cdots\label{eq:ascaledh}\\
\frac{db}{d\tau}&=&b+{\rm Re}[c(k_2,v_2+i0)]\,b^3
+\gamma^{4}\,{\rm Re} [e^{i(\theta_b+\theta_a)}(Q_1(0)+P_3(0))]a b^2
+\cdots.\label{eq:bscaledh}
\end{eqnarray}
The joint conditions $k_2+k_1=0$ and $v_2=-v_1$ characteristic of this
instability imply $\omega_2=\omega_1$; hence the exponentials in the
mode coupling terms are rapidly oscillating and we expect these terms to
have a negligible effect on the time-averaged evolution. In effect, the coupled
system (\ref{eq:ascaledh}) - (\ref{eq:bscaledh}) reduces to
(\ref{eq:ascalednr2}) - (\ref{eq:bscalednr2}) and we recover the
superposition picture of the two single mode amplitude equations
but without the resonant $a^3$ term.

If the ions are taken to be infinitely massive, then Table 3 replaces Table 2
and the scaling associated with the single mode instabilities becomes
$\beta_j=2$.  The re-analysis of (\ref{eq:ascaled}) - (\ref{eq:bscaled})
leads similar conclusions although with quantitative differences. Only for the
resonance $k_2=2k_1$ and $v_2=v_1$ are the single mode scalings altered and
again the short wavelength mode is suppressed; in this case we find $\beta_2$
increases to $\beta_2=5/2$. When $v_2\neq v_1$ we again find that the amplitude
equations decouple as in (\ref{eq:ascalednr2}) - (\ref{eq:bscalednr2}).

\section{Discussion}

In this paper we have investigated the interaction between two weakly unstable
electrostatic waves in an unmagnetized plasma. Such growing modes are defined
unambiguously by eigenfunctions of the linear theory;
Landau-damped collective modes are properly thought of as part of the continuum
and hence are not decaying``modes'' in the present sense.\cite{cra1} For the
growing modes
discussed here, the theory provides a self-consistent description that includes
the nonlinear effects of wave-particle resonance.
Such a formulation, while more complex,
captures phenomena that are absent from conventional formulations involving
wave interactions.\cite{pc78}

The wave-wave interaction is described by
coupled amplitude equations consistent with the assumed translation invariance
of the system and any symmetry of the unstable velocity distribution function.
The coupling coefficients exhibit new singularities in the weakly unstable
limit, but except in the case of the two-to-one spatial resonance do not
alter the scaling for the saturation amplitude identified in single
mode theory, at least through third order in the mode amplitudes. In the
special but important case of two-to-one spatial resonance the overlap of
velocity resonances and spatial resonance modifies the scaling, resulting in a
dramatic suppression of the short wave mode.

In the absence of velocity overlap and the two-to-one spatial
resonance the single mode scaling shows that the two waves evolve as a
simple superposition of the individual instabilities.
This picture is qualitatively consistent
with the numerical results obtained
by Demeio and Zweifel for beam-plasma instabilities with reflection
symmetry.\cite{demeio} In addition, it agrees more quantitatively with the
analysis of Buchanan and Dorning
 who constructed superpositions of BGK modes
as candidates for the asymptotic states produced in the numerical
simulations.\cite{bd93} In particular, these authors found that
a consistent construction to leading order in the amplitude of the
individual BGK modes required unequal phase velocities;
an assumption analogous to our $v_2\neq v_1$ condition.
Our results suggest a precise connection between the initial value problem
for the unstable waves and the superposed
BGK states of Buchanan and Dorning.

The significance of resonance overlap, such as $v_2=v_1$,
in the single particle
phase space is well-established from studies of particle motion
in fields produced by large amplitude waves. In these studies
the appearance of chaotic particle trajectories
is investigated, but the
self-consistent modification of the wave evolution by the particles is
routinely neglected.\cite{chir79,ll83,esc85}
The situation we consider does not allow this simplifying
approximation since the resonant particles drive the linear instability
and also dominate the nonlinear evolution of the waves.
It is striking to discover
that resonance ovelap has a profound effect on the dynamics of the waves,
in addition to its better known consequences for the associated particle
dynamics.

Our prediction that resonant interaction with a longer wavelength
mode can modify the nonlinear scaling of a short wavelength mode
may be amenable to experimental test. The single mode scaling
for fixed ions was detected experimentally in  measurements
on an electron beam injected through a travelling wave tube.\cite{tdm87}
The tube plays the role of the non-resonance
electrons and supports a propagating wave which couples to the
resonant particles in the beam.
If the electron beam is sufficiently cold the unstable
waves approximately satisfy the approximate dispersion relation
$\omega(k)=v_p k$ with constant $v_p$ and hence always have equal
phase velocities.\cite{jdc95b}
Under these conditions, one
should measure the scaling of a single mode launched at
$k_2$ and then repeat the measurement when a second wave is launched
simultaneously at $k_1=k_2/2$.

The possibility remains that singularities
in the higher order coupling coefficients could modify these conclusions and
force new scalings for other spatial resonances as well. However, in the
study of singular amplitude equations in other problems it is commonly
found that the dominant singularities appear in the low order nonlinear
terms.\cite{jdcaj97,jdctd97}
In addition, the fact that the $k_2=3k_1$ spatial resonance does
not shift the scalings may signify that resonances other than $k_2=2k_1$
will generally leave the single mode scalings undisturbed. These speculations
can be tested by examining the singularities in the amplitude expansions
to all orders. Such an investigation may be feasible using the
techniques of (I).

\section{\hspace{0.125in}Acknowledgements}

This work supported by NSF grant PHY-9423583.


\clearpage


\begin{thebibliography}{99}
\bibitem{relprime} We assume that $n_1$ and $n_2$ have no common integer
divisors other than unity;
if this is not the case then the integers used in
(\ref{eq:adef}) - (\ref{eq:invariants})
are obtained from $n_1$ and $n_2$ by factoring out the common divisor.
\end{thebibliography}
\end{document}